\documentclass[journal=apchd5,manuscript=article,]{achemso}
\usepackage{amsmath,amssymb}
\usepackage[version=3]{mhchem} % Formula subscripts using \ce{}
\usepackage{orcidlink}
\usepackage{pdfpages}
\hypersetup{hidelinks}

\newcommand{\ohba}{\textit{o}HBA}
\newcommand{\od}{O$_\mathrm{D}$}
\newcommand{\oa}{O$_\mathrm{A}$}
\newcommand*\revision[1]{\textcolor{black}{#1}}
\author{Jonathan H. Fetherolf\orcidlink{0000-0001-6470-1461}}
\affiliation{Department of Chemistry, Princeton University, Princeton, NJ 08544, USA}
\author{Tim Duong\orcidlink{0009-0001-7551-8504}}
\affiliation{Department of Chemistry, Princeton University, Princeton, NJ 08544, USA}
\author{Tao E. Li\orcidlink{0000-0003-0361-5687}}
\affiliation{Department of Physics and Astronomy, University of Delaware, Newark, DE 19716, USA}

\author{Sharon Hammes-Schiffer\orcidlink{0000-0002-3782-6995}}
\email{shs566@princeton.edu}
\affiliation{Department of Chemistry, Princeton University, Princeton, NJ 08544, USA}

\title
  {Nuclear–Electronic Quantum Dynamics in a 
  Plasmonic Nanocavity}

\let\oldmaketitle\maketitle
\let\maketitle\relax

\begin{document}

%\twocolumn[
\begin{@twocolumnfalse}
\oldmaketitle
\begin{abstract}
Plasmonic nanocavities are a promising platform for strong light--matter coupling and enhanced spectroscopies at the single-molecule level. These nanoscale environments are challenging to model due to their strongly multimodal character and short cavity lifetimes. Herein, we study the effects of these environments using real-time nuclear--electronic orbital time-dependent density functional theory (RT-NEO-TDDFT) coupled to multiple classical cavity modes in a manner that includes cavity loss. In RT-NEO-TDDFT, the quantum mechanical densities of all electrons and specified nuclei, typically protons, are propagated in real time. We show that a cavity with many modes at different frequencies can be used to probe and modify the nuclear--electronic quantum dynamics of chemical systems. Ultrafast excited-state proton transfer reactions can be probed through the time- and energy-resolved cavity emission of a multimode cavity. Under strong coupling conditions, the cavity can modify the dynamics, in some cases suppressing proton transfer and exhibiting Rabi-like oscillations of the cavity emission due to polariton formation. Utilizing the spectral density for an experimentally relevant nanoparticle-on-mirror single-molecule cavity, we show that an excited-state proton transfer system can evolve into resonance with the cavity even when initially out of resonance with the dominant cavity peak. In this case, tuning the dominant cavity peak to be resonant with the electronic transition leads to polariton formation for a small collection of molecules. The RT-NEO framework with multimode cavities enables the efficient simulation of chemical reactions in physically realistic electromagnetic environments, providing fundamental insights into the dynamics and associated spectroscopic signatures. 

\vspace{1.0em}
\end{abstract}
\end{@twocolumnfalse}
%]

\section*{Introduction}

Plasmonic nanocavities are nanoscale optical antennae configured to achieve extreme electric field enhancement in a confined region of space.\cite{novotny_vanhulst_2011,giannini_fernandez-dominguez_heck_maier_2011} Because they rely on localized surface plasmon resonances, rather than standing electromagnetic waves as in micron-scale Fabry-Perot optical cavities,\cite{yokoyama_1992} plasmonic nanocavities are able to concentrate fields into dimensions much smaller than the resonant wavelength. This nanoscale localization has allowed the experimental realization of polaritons, hybrid light--matter states achieved through strong coupling, on the few- or even single-molecule scale.\cite{Zengin_Wersall_Nilsson_Antosiewicz_etal_2015,Chikkaraddy_deNijs_Benz_Barrow_Scherman_Rosta_etal_2016, Santhosh_Bitton_Chuntonov_Haran_2016} In addition to the strong-coupling regime, nanocavities have been widely used for a variety of enhanced spectroscopy techniques, including single-molecule detection and other nanoscale sensing applications.\cite{Nie_Emory_1997,Kneipp_Wang_Kneipp_Perelman_etal_1997,Kinkhabwala_Yu_Fan_Avlasevich_etal_2009,Benz_Schmidt_Dreismann_Chikkaraddy_etal_2016,Maccaferri_Barbillon_NanaKoya_Lu_etal_2021} These advances underscore the need for theoretical treatments that faithfully capture the underlying microscopic processes taking place in both the intermediate- and strong-coupling regimes.

Molecular polaritons have garnered widespread theoretical and experimental interest in recent years due to their potential for controlling molecular properties via hybridization with light.\cite{Basov_Asenjo-Garcia_Schuck_Zhu_etal_2020,Garcia-Vidal_Ciuti_Ebbesen_2021,Xiang_Xiong_2024,Li_Cui_Subotnik_Nitzan_2022,Mandal_Taylor_Weight_Koessler_Li_Huo_2023,Ebbesen_Rubio_Scholes_2023} For example, experimental evidence suggests modified thermal\cite{Thomas_George_Shalabney_Dryzhakov_etal_2016,Thomas_Lethuillier-Karl_Nagarajan_Vergauwe_etal_2019,Ahn_Triana_Recabal_Herrera_etal_2023} and photochemical\cite{Hutchison_Schwartz_Genet_Devaux_etal_2012, Ng_Dligatch_Amekura_Davis_etal_2015, Munkhbat_Wersall_Baranov_Antosiewicz_Shegai_2018, Satapathy_Khatoniar_Parappuram_Liu_etal_2021} reaction rates, as well as enhanced charge carrier\cite{Orgiu_George_Hutchison_Devaux_Dayen_Doudin_etal_2015,Bhatt_Kaur_George_2021, Nagarajan_George_Thomas_Devaux_Chervy_Azzini_etal_2020} and exciton\cite{Liu_Huang_Hou_Fan_Forrest_2022,Liu_Lynch_Zhao_Conran_McAleese_Jariwala_etal_2023,Xu_Mandal_Baxter_Cheng_Lee_Su_etal_2023} transport under strong coupling. The vast majority of observed effects take place in the collective regime, in which the light--matter coupling is strengthened via interaction with a relatively large ensemble of molecules. Specifically, when $N$ two-level emitters are coupled to an electromagnetic cavity mode as in the Tavis-Cummings model\cite{Tavis_Cummings_1968,Tavis_Cummings_1969}, the light--matter coupling scales as $\sqrt{N/V}$, where $V$ is the effective cavity volume. Based on this general scaling and the relative ease of procurement and manipulation of micron-scale optical cavity setups, it is therefore unsurprising that realizations of molecular polaritons have been largely in this regime. This $\sqrt{N/V}$ scaling also demonstrates that, given a small enough cavity volume $V$, such as those in the sub-wavelength regime of plasmonic nanocavities, strong coupling can be achieved with only a few or even single molecules. However, experimental and theoretical studies of nanocavity setups are complicated by significantly shorter cavity lifetimes (i.e., they are much more lossy) than micron-scale optical cavities, and the cavity field (i.e., the spectral density) is strongly multimodal in character.\cite{Khurgin_2015, Matsuzaki_Liu_Gotzinger_Sandoghdar_2021, Fregoni_Garcia-Vidal_Feist_2022} Although this more complex, lossy electromagnetic environment is generally viewed as a detriment to practical applications, several studies have demonstrated that cavity loss can be exploited to channel energy away from unwanted reaction pathways such as photo-oxidation.\cite{Galego_Garcia-Vidal_Feist_2016,Munkhbat_Wersall_Baranov_Antosiewicz_Shegai_2018,Felicetti_Fregoni_Schnappinger_Reiter_etal_2020,Torres-Sanchez_Feist_2021,Antoniou_Suchanek_Varner_Foley_2020} Additionally, the associated cavity emission could potentially serve as an ultrafast probe of excited-state molecular dynamics\cite{Silva_Pino_Garcia-Vidal_Feist_2020}, a concept we will explore further in this work.

Theoretical modeling of molecules coupled to electromagnetic environments is challenging due to the potentially vast number of electronic, nuclear, and cavity mode degrees of freedom. Because of this complexity, approximations must be made in some part of the problem to reduce its dimensionality. Quantum electrodynamical (QED) electronic structure theory treats the molecule, electromagnetic field, and their interaction fully quantum mechanically within the nuclear Born--Oppenheimer approximation.\cite{Ruggenthaler_Sidler_Rubio_2023,Foley_McTague_DePrince_2023,Ruggenthaler_Mackenroth_Bauer_2011,Ruggenthaler_Flick_Pellegrini_Appel_etal_2014,Haugland_Schafer_Ronca_Rubio_etal_2021,Yang_Ou_Pei_Wang_etal_2021,Pavosevic_Rubio_2022,Vu_Mejia-Rodriguez_Bauman_Panyala_etal_2024,Weight_Tretiak_Zhang_2024,Weber_dosAnjosCunha_Morales_Rubio_etal_2025,Tasci_Cunha_Flick_2025,Castagnola_Riso_ElMoutaoukal_Ronca_etal_2025,DePrince_2021,Mallory_DePrince_2022,ElMoutaoukal_Riso_Castagnola_Ronca_etal_2025,Bauer_Dreuw_2023,Cui_Mandal_Reichman_2024,Flick_Welakuh_Ruggenthaler_Appel_Rubio_2019,Flick_Ruggenthaler_Appel_Rubio_2015,Flick_Narang_2020} While QED methods capture the electron--photon interaction with a high degree of accuracy, they are generally limited to small system sizes and simple, lossless electromagnetic spectral structures, in addition to neglecting nuclear--electronic nonadiabatic effects. QED electronic structure methods have been implemented with multimode cavity setups, but such treatments are challenging due to the fully quantum mechanical nature of the light--matter interaction.\cite{Tichauer_Feist_Groenhof_2021,Wang_Neuman_Flick_Narang_2021,Svendsen_Thygesen_Rubio_Flick_2024} Other approaches have utilized nonadiabatic dynamical methods for the electron--photon or nuclear--electronic interactions, for example via Ehrenfest\cite{Hoffmann_Schafer_Rubio_Kelly_Appel_2019,Hoffmann_Lacombe_Rubio_Maitra_2020,Rosenzweig_Hoffmann_Lacombe_Maitra_2022,Zhou_Hu_Mandal_Huo_2022}, surface hopping,\cite{Martinez_Rosenzweig_Hoffmann_Lacombe_Maitra_2021} or ab initio multiple spawning\cite{Curchod_Martinez_2018} approaches. Several approaches have also been developed for accurately describing the multimode nanocavity environment via a spectral density based on solving the classical electrodynamics equations. \cite{Cuartero-Gonzalez_Fernandez-Dominguez_2018,Cuartero-Gonzalez_2020,Medina_Garcia-Vidal_Fernandez-Dominguez_Feist_2021,Sanchez-Barquilla_Garcia-Vidal_Fernandez-Dominguez_Feist_2022} However, few methods exist that accurately capture the nuclear--electronic quantum dynamics of molecules interacting strongly with complex electromagnetic environments.

The nuclear--electronic orbital (NEO) approach is a powerful quantum chemistry framework in which select nuclei, typically protons, are treated at the same level of theory as electrons.\cite{Webb_Iordanov_Hammes-Schiffer_2002,Pavosevic_Culpitt_Hammes-Schiffer_2020,Hammes-Schiffer_2021} Real-time NEO time-dependent density functional theory (RT-NEO-TDDFT)\cite{Zhao_Tao_Pavosevic_Wildman_etal_2020} has been used to model polaritons in both the vibrational and electronic strong coupling regimes.\cite{Li_Tao_Hammes-Schiffer_2022} In most implementations, the cavity modes are treated as classical harmonic oscillators, an approach adapted from the fully classical CavMD method for vibrational strong coupling.\cite{Li_Subotnik_Nitzan_2020,Li_Nitzan_Subotnik_2021} Recently, RT-NEO-TDDFT was implemented with a fully quantum mechanical treatment of the cavity degrees of freedom,  demonstrating that the classical and quantum  treatments give very similar results for many observables of interest.\cite{Welman_Li_Hammes-Schiffer_2025} Another recent study used NEO time-dependent configuration interaction\cite{Garner_Upadhyay_Li_Hammes-Schiffer_2024} and classical cavity modes to study polaritons generated from mixed nuclear--electronic vibronic excitations as well as hydrogen tunneling splittings.\cite{Garner_Li_Hammes-Schiffer_2025} Real-time NEO methods provide the quantum mechanical time evolution of key nuclei, such as transferring protons, as well as the electrons, without invoking the Born--Oppenheimer approximation between them. Within the classical cavity mode approximation, the cost does not substantially increase with additional electromagnetic degrees of freedom. For this reason, RT-NEO methods with classical cavity modes are an excellent platform for studying the effects of multimode electromagnetic environments on nuclear--electronic quantum dynamics.

Herein, we explore the use of a multimode, lossy cavity as a time-resolved spectroscopic probe of nuclear--electronic quantum dynamics. Our approach is inspired by the work of Silva and coworkers\cite{Silva_Pino_Garcia-Vidal_Feist_2020}, who used the emission from a single cavity mode to track wave packet evolution on an electronic excited-state surface upon off-resonant pumping. We propose that a multimode cavity environment can be used to energetically resolve the nuclear--electronic dynamics due to differential emission from modes across the intrinsically broadened spectral region. In the intermediate-coupling regime, where the molecular properties are not modified by the cavity, this treatment could provide a powerful time-resolved single-molecule sensing technique. In the strong-coupling regime, this treatment can be used to study the dynamical process of light--matter hybridization in polariton formation and the impact of chemical reactions such as proton transfer. Within this theoretical framework, we also study how the geometric and material parameters of the nanocavity can be tuned to achieve single- or few-molecule strong coupling.

\section*{Theory}
We begin with the full QED Hamiltonian,
\begin{equation}
    \hat{H}_\text{QED}=\hat{H}_\text{M}+\hat{H}_\text{F}
\end{equation}
where $\hat{H}_\text{M}$ is the molecular Hamiltonian and $\hat{H}_\text{F}$ describes the electromagnetic field. The molecular Hamiltonian is
\begin{equation}
    \hat{H}_\text{M}=\sum_i \bigg[ \frac{\hat{\mathbf{p}}_i^2}{2m_i} + \hat{V}(\{\hat{\mathbf{r}}_i\},\mathbf{R}_\text{c}) \bigg]
\end{equation}
where $\hat{\mathbf{p}}_i$, $m_i$ and $\hat{\mathbf{r}}_i$ are the momenta, masses,  and positions of the quantum particles (in this case, electrons and protons), and $\hat{V}$ is the Coulomb potential between all particles, including the classical nuclear coordinates $\mathbf{R}_\text{c}$, which are assumed to be fixed in this work. The field Hamiltonian can be written as\cite{Wang_Neuman_Flick_Narang_2021} 
\begin{equation}
    \hat{H}_\text{F}=\frac{1}{2}\sum_{\alpha} \bigg[ \hat{p}^2_\alpha + \omega_\alpha^2\bigg( \hat{q}_\alpha + \frac{\lambda_\alpha}{\omega_\alpha}\hat{\boldsymbol{\mu}}_\text{M}\cdot\hat{\boldsymbol{\xi}}_s \bigg)^2\bigg].
\end{equation}
Here, $\alpha$ is the compound mode index $\alpha=(k,s)$, where $k=|\mathbf{k}|$ is the mode-specific wave vector magnitude and $s$ is the polarization direction. Each mode has a momentum $\hat{p}_\alpha$, displacement $\hat{q}_\alpha$, and frequency $\omega_\alpha$. Moreover, $\hat{\boldsymbol{\mu}}_\text{M}$ is the molecular dipole operator, and $\hat{\boldsymbol{\xi}}_s$ is the unit vector in the $s$ direction. 

The cavity field strength $\lambda_\alpha$ is given as 
\begin{equation}
    \lambda_\alpha=e\sqrt{\frac{2}{\hbar\omega_\alpha}}E_\alpha
\end{equation}
where $e$ is the elementary unit of charge and $E_\alpha$ is the electric field amplitude at the center of charge density. The cavity field strength is related to the light--matter coupling via the spectral density as
\begin{equation}
     \mu_s^2\lambda_\alpha^2 = \int_0^\infty d\omega J_\alpha(\omega)
\end{equation}
where $J_\alpha(\omega)$ is the Lorentzian spectral density describing the lineshape of mode $\alpha$ due to finite cavity lifetime
\begin{equation}
J_\alpha(\omega)=\frac{g_\alpha^2}{\pi}\frac{\gamma_\alpha/2}{(\omega-\omega_\alpha)^2+(\gamma_\alpha/2)^2}.
\label{eq:lorentzian}
\end{equation}
Here, $g_\alpha$ is the light--matter coupling for mode $\alpha$, and $\gamma_\alpha$ is the linewidth of the peak centered at $\omega_\alpha$.

Neglecting the self-dipole term and adopting the semiclassical approximation, $\hat{H}_\text{QED}$ becomes
\begin{equation}
    \hat{H}_\text{QED} = \hat{H}_\text{sc}+\frac{1}{2}\sum_\alpha \bigg[ p^2_\alpha + \omega^2_\alpha q^2_\alpha \bigg]
\end{equation}
where the semiclassical light--matter Hamiltonian is 
\begin{equation}
    \hat{H}_\text{sc} = \hat{H}_\text{M} + \sum_\alpha \varepsilon_\alpha q_\alpha \hat{\mu}_s.
    \label{eq:ham_sc}
\end{equation}
The dipole component in the $s$ direction is $\hat{\mu}_s\equiv\hat{\boldsymbol{\mu}}_\text{M}\cdot\hat{\boldsymbol{\xi}}_s$, and $\varepsilon_\alpha\equiv\omega_\alpha \lambda_\alpha$ is the semiclassical cavity field strength for each mode $\alpha$. The field strength $\varepsilon_\alpha$ can either be a free parameter that is tuned to set the total coupling strength or be derived by solving the classical electrodynamics equation for a nanocavity setup.\cite{Cuartero-Gonzalez_Fernandez-Dominguez_2018,Cuartero-Gonzalez_2020,Svendsen_Thygesen_Rubio_Flick_2024} \revision{The effect of the self-dipole term, i.e., the term proportional to $\lambda_\alpha^2\mu_s^2$, has been demonstrated to mainly be confined to the ultrastrong coupling regime.\cite{Schafer_Ruggenthaler_Rokaj_Rubio_2020,Flick_Ruggenthaler_Appel_Rubio_2017a} Because the systems under study are well outside this regime, we neglect this term as has been done in previous studies using this same Hamiltonian.\cite{Li_Tao_Hammes-Schiffer_2022,Welman_Li_Hammes-Schiffer_2025,Garner_Li_Hammes-Schiffer_2025} However, future work applying the RT-NEO method to systems with larger coupling will include the self-dipole term.
}

Each cavity mode is propagated via the classical equations of motion: 
\begin{subequations}
    \begin{equation}
        \dot{q}_\alpha=p_\alpha
    \end{equation}
    and
    \begin{equation}
        \dot{p}_\alpha=-\omega_\alpha^2 q_\alpha - \varepsilon_\alpha\mu_s - \gamma_\alpha p_\alpha
    \end{equation}
\end{subequations}
where $p_\alpha$ and $q_\alpha$ are the classical momentum and position displacement of mode $\alpha$. The loss term with mode-specific inverse lifetime (loss rate) $\gamma_\alpha$ corresponds to the Lorentzian linewidth in the spectral density. The magnitude of the total molecular dipole in the $s$ direction $\mu_s$ is computed as the expectation value at time $t$ minus the permanent value at time $t=0$, $\mu_s=\langle\mu_s(t)\rangle-\langle\mu_s(t=0)\rangle$, to avoid spurious mode excitations at $t=0$ as a result of the static dipole. 

For each mode, the far-field cavity emission rate $I_\alpha$ is proportional to the mode-specific loss rate $\gamma_\alpha$, as well as the occupation number expectation value for that mode:
\begin{equation}
    I_\alpha = \gamma_\alpha\langle \hat{a}_\alpha^\dagger \hat{a}_\alpha \rangle.
\end{equation}
Here, $\hat{a}_\alpha$ and $\hat{a}_\alpha^\dagger$ are the harmonic oscillator ladder operators associated with mode $\alpha$, and $\hat{a}_\alpha^\dagger \hat{a}_\alpha$ is the number operator.  The number operator is related to the momentum and position via the harmonic oscillator energy expression, $\hbar\omega_\alpha \hat{a}_\alpha^\dagger \hat{a}_\alpha = \frac{1}{2}(\hat{p}^2_\alpha + \omega^2_\alpha \hat{q}^2_\alpha$). Making the semiclassical approximation, this expression allows us to determine the instantaneous emission rate at time $t$ in terms of the classical momentum and displacement of each mode, $p_\alpha(t)$ and $q_\alpha(t)$:
\begin{equation}
    I_\alpha(t) = \frac{\gamma_\alpha}{2\hbar\omega_\alpha}\bigg( p^2_\alpha(t) + \omega^2_\alpha q^2_\alpha(t) \bigg).
    \label{eq:emission}
\end{equation}
We will henceforth refer to emission rate simply as emission.
%and the time-resolved emission spectrum can be obtained by summing over modes:
%\begin{equation}
%    I(\omega,t)=\sum_\alpha I_\alpha(t) \delta(\omega-\omega_\alpha).
%\end{equation}

The subsystem containing the molecule under the influence of the classical cavity modes is propagated via the time-dependent Schr\"odinger equation,
\begin{equation}
    i\hbar \frac{\partial}{\partial t}\Psi(\mathbf{x}^\text{e},\mathbf{x}^\text{n}; t) = \hat{H}_\text{sc}(\mathbf{x}^\text{e},\mathbf{x}^\text{n}; t)\Psi(\mathbf{x}^\text{e},\mathbf{x}^\text{n}; t)
\end{equation}
where $\Psi$ is the nuclear--electronic wavefunction, and $\mathbf{x}^\text{e}$ and $\mathbf{x}^\text{n}$ are the spatial and spin coordinates for the electronic and quantum nuclear (protonic) degrees of freedom. The electronic and quantum nuclear density matrices are propagated via the von Neumann equations:
\begin{subequations}
    \begin{equation}
        i\hbar\frac{\partial}{\partial t} \mathbf{P}^\text{e}(t) = \bigg[ \mathbf{F}^\text{e}(t) + \sum_\alpha \varepsilon_\alpha q_\alpha(t)\hat{\mu}^\text{e}_s, \mathbf{P}^\text{e}(t) \bigg]
    \end{equation}
and
    \begin{equation}
        i\hbar\frac{\partial}{\partial t} \mathbf{P}^\text{n}(t) = \bigg[ \mathbf{F}^\text{n}(t) + \sum_\alpha \varepsilon_\alpha q_\alpha(t)\hat{\mu}^\text{n}_s, \mathbf{P}^\text{n}(t) \bigg]
    \end{equation}
\end{subequations}
where $\mathbf{P}^\text{e}$ and $\mathbf{P}^\text{n}$ are the electronic and protonic density matrices, which are initially constructed from their coefficient matrices $\mathbf{C}^\text{e}$ and $\mathbf{C}^\text{n}$, respectively, in the orthonormal atomic orbital basis. Furthermore, $\hat{\mu}^\text{e}_s$ and $\hat{\mu}^\text{n}_s$ are the electronic and quantum nuclear dipole moments, respectively, in the $s$ direction, where $s$ indicates the $x$, $y$ or $z$ direction. 

In these equations, $\mathbf{F}^\text{e}$ and $\mathbf{F}^\text{n}$ are the electronic and protonic Kohn-Sham matrices:
\begin{subequations}
\begin{equation}
\begin{split}
    \mathbf{F}^\text{e}(t) &= \mathbf{H}^\text{e}_\text{core} + \mathbf{J}^\text{ee}(\mathbf{P}^\text{e}(t)) + \mathbf{V}^\text{e}_\text{xc}(\mathbf{P}^\text{e}(t)) \\
    &- \mathbf{J}^\text{en}(\mathbf{P}^\text{n}(t)) + \mathbf{V}^\text{en}_\text{c}(\mathbf{P}^\text{e}(t), \mathbf{P}^\text{n}(t))% + \mathbf{V}^\text{e}_\text{ext}(t)
\end{split}
\end{equation}
and
\begin{equation}
\begin{split}
    \mathbf{F}^\text{n}(t) &= \mathbf{H}^\text{n}_\text{core} + \mathbf{J}^\text{nn}(\mathbf{P}^\text{n}(t)) + \mathbf{V}^\text{n}_\text{xc}(\mathbf{P}^\text{n}(t))\\ 
    &- \mathbf{J}^\text{ne}(\mathbf{P}^\text{e}(t)) + \mathbf{V}^\text{ne}_\text{c}(\mathbf{P}^\text{n}(t), \mathbf{P}^\text{e}(t)).% + \mathbf{V}^\text{n}_\text{ext}(t).
\end{split}
\end{equation}
\end{subequations}
Here, $\mathbf{H}_\text{core}^\text{e(n)}$ is the core Hamiltonian that includes the kinetic energy of the electrons (quantum nuclei) and the Coulomb interaction between the electrons (quantum nuclei) and the classical nuclei. $\mathbf{J}^\text{ee(nn)}$ is the Coulomb repulsion between electrons (quantum nuclei), while $\mathbf{V}_\text{xc}^\text{e(n)}$ is the exchange-correlation potential for electrons (quantum nuclei). $\mathbf{J}^\text{en(ne)}$ is the attractive electronic-quantum nuclear Coulomb interaction, and $\mathbf{V}_c^\text{en(ne)}$ is the electron-proton correlation potential. 
%Finally, $\mathbf{V}_\text{ext}^\text{e(n)}$ is the external potential
%
\section*{Simulation details}
All computations were performed in a development branch of the Q-Chem software package.\cite{Epifanovsky_Gilbert_Feng_Lee_Mao_Mardirossian_etal_2021} Molecular structures were optimized using conventional DFT for the ground state and TDDFT for the excited states. For all NEO calculations, three basis function centers were used for the transferring proton: one was optimized near the donor in the ground state (i.e., the optimized ground-state position), one was optimized near the acceptor in the S$_1$ state, and one was placed at the midpoint between these two positions. All calculations were performed using the cc-pVDZ electronic basis set\cite{Dunning_1989} and the B3LYP electronic exchange-correlation functional\cite{Lee_Yang_Parr_1988,Becke_1993}. NEO calculations were performed with the PB4D protonic basis set\cite{Yu_Pavosevic_Hammes-Schiffer_2020} and the epc17-2 electron-proton correlation functional.\cite{Yang_Brorsen_Culpitt_Pak_Hammes-Schiffer_2017,Brorsen_Yang_Hammes-Schiffer_2017} For all cavity calculations, the polarization direction was aligned with the S$_0$ to S$_1$ transition dipole vector. The classical cavity modes were propagated using the velocity Verlet algorithm. The quantum molecular equations of motion were propagated via a modified midpoint unitary transform time-propagation scheme\cite{Li_Smith_Markevitch_Romanov_etal_2005,Goings_Lestrange_Li_2018} with an additional predictor--corrector step to control numerical error.\cite{DeSantis_Storchi_Belpassi_Quiney_etal_2020} A timestep of 0.1 a.u. was used for all real-time calculations. \revision{All parameters used for generating the cavity spectral densities are given in Tables S1 and S2 of the Supporting Information.}
\section*{Results and discussion}
\subsection*{Excited-state proton transfer in a Gaussian multimode environment}
\begin{figure*}[ht]
    \centering
    \includegraphics[width=5in]{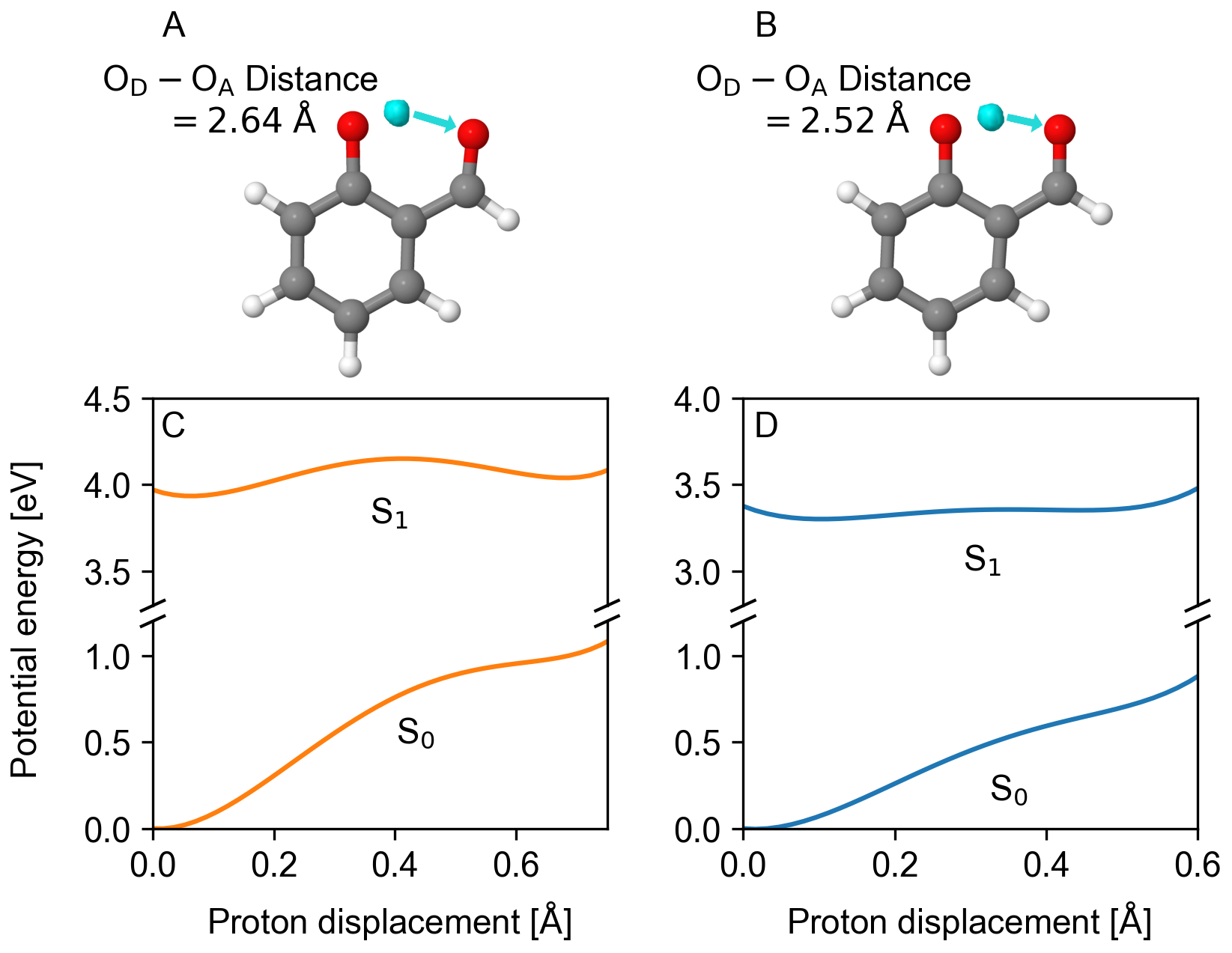}
    \caption{Upper portion: Structure of \ohba~at (A) the conventional DFT optimized ground-state (S$_0$) geometry and (B) the conventional TDDFT S$_1$ excited-state geometry with the transferring proton constrained to have the same distance from the donor oxygen O$_\mathrm{D}$ as in the ground state. The isosurface of the transferring proton density computed with NEO-DFT is shown in cyan. The cyan arrow depicts the proton transfer coordinate used for computing the potential energy profiles. Lower plots: Proton potential energy profiles for the S$_0$ and S$_1$ states in the (C) S$_0$ geometry and (D) constrained S$_1$ geometry, as computed with conventional TDDFT by scanning the proton coordinate along the axis connecting its optimized ground state position to O$_\text{A}$.}
    \label{fig:ohba_pes}
\end{figure*}

We apply our approach for modeling the nuclear--electronic quantum dynamics in a multimode cavity to the excited-state intramolecular proton transfer (ESIPT) reaction of \textit{o}-hydroxybenzaldehyde (\ohba). This reaction has been widely studied both experimentally\cite{Lochbrunner_Schultz_Schmitt_Shaffer_etal_2001,Stock_Bizjak_Lochbrunner_2002} and theoretically.\cite{Aquino_Lischka_Hattig_2005,Scheiner_2000,Zhao_Tao_Pavosevic_Wildman_etal_2020,Zhao_Wildman_Pavosevic_Tully_etal_2021,Li_Tao_Hammes-Schiffer_2022,Chow_Li_Hammes-Schiffer_2023,Xu_Zhou_Blum_Li_etal_2023a}. The structure of \ohba~is shown on the top of Fig. \ref{fig:ohba_pes}, with the transferring proton represented in cyan. Proton transfer occurs upon excitation from the S$_0$ ground state to the S$_1$ excited state via a $\pi$ to $\pi^*$ transition. Previous studies with the RT-NEO-TDDFT method demonstrated that quantization of the transferring proton leads to an accelerated rate, which occurs on the ultrafast timescale ($\approx$10-100 fs).\cite{Zhao_Tao_Pavosevic_Wildman_etal_2020,Zhao_Wildman_Pavosevic_Tully_etal_2021} The occurrence and timescale of the proton transfer depends strongly on geometric fluctuations of the molecular backbone, and in particular the distance between the proton donor oxygen \od~and acceptor oxygen \oa. 

In this work, we use two fixed heavy-atom geometries, one that prevents proton transfer and one that allows proton transfer. These geometries and the corresponding proton potential energy profiles are shown in Fig. \ref{fig:ohba_pes}. These proton potentials were obtained with conventional TDDFT by moving the proton along the axis connecting its optimized ground state position to O$_\text{A}$. The structure in Fig. \ref{fig:ohba_pes}A and potential in Fig. \ref{fig:ohba_pes}C correspond to the ground-state geometry, where the \od$-$\oa~distance is 2.64 \AA. Figs. \ref{fig:ohba_pes}B and \ref{fig:ohba_pes}D show the structure and potential corresponding to the geometry optimized on the S$_1$ surface, where the optimization was performed with a constraint on the \od$-$H distance to maintain the ground-state bond length. This geometry has a smaller \od$-$\oa~distance of 2.52 \AA, and the corresponding S$_1$ proton potential energy profile is both lower in energy overall and has a much lower barrier between the proton donor and acceptor minima.

\begin{figure*}[htp]
    \centering
    \includegraphics[width=6in]{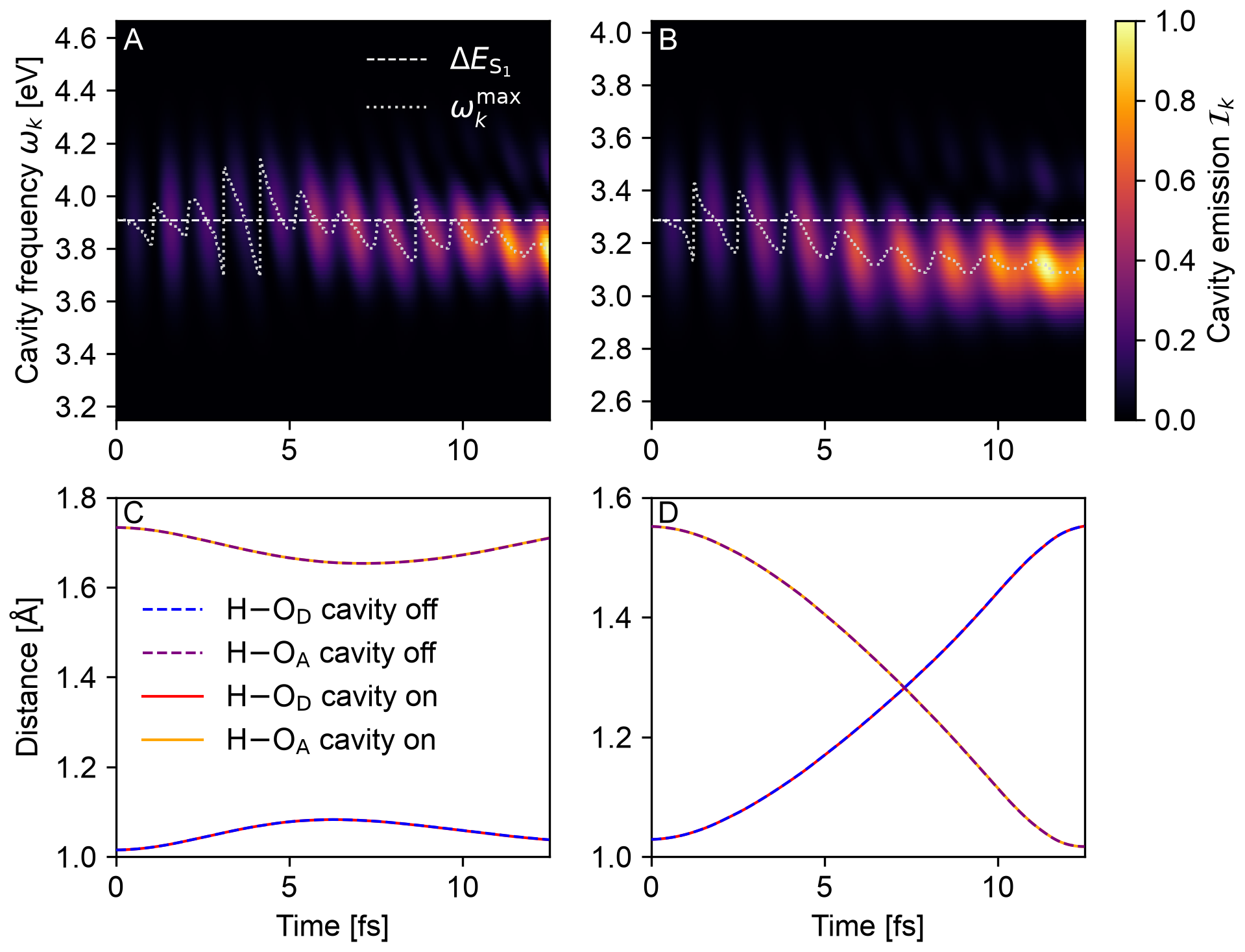}
    \caption{Upper plots: Mode-specific cavity emission, $I_k$, indicated by color intensity, as a function of mode frequency $\omega_k$ and time, for \ohba~in the (A) S$_0$ geometry and (B) constrained S$_1$ geometry. The cavity environment is represented by a Gaussian spectral density peaked at the initial NEO-TDDFT vertical S$_0$ to S$_1$ excitation energy, $\Delta E_{\text{S}_1}$, with $g_0/\mu_0=0.03$ V/nm and $\sigma=0.25$ eV. The frequency of the cavity mode with the largest emission as a function of time is $\omega^\text{max}_k$. Emission is normalized by the single largest intensity observed across the entire time series.  Lower plots: H--\od~and H--\oa~distances in \AA ngstroms as a function of time for \ohba~in the (C) S$_0$ geometry and (D) constrained S$_1$ geometry. The distances are computed using the proton position expectation value. The results are shown with the cavity off \revision{(blue and violet dashed) and on (red and orange solid)}. Note that, in this case, the cavity has no effect on the proton transfer dynamics, and therefore the results with the cavity on and off are indistinguishable.}
    \label{fig:4panel_weak}
\end{figure*}

We first establish the bare, out-of-cavity proton transfer dynamics of \ohba~using RT-NEO-TDDFT. All ESIPT calculations are initialized with the NEO-DFT closed-shell restricted ground-state density. To model photoexcitation for all ESIPT calculations, one electron is promoted from the highest occupied molecular orbital (HOMO) to the lowest unoccupied molecular orbital (LUMO), and the unrestricted open-shell singlet density is propagated. The HOMO--LUMO transition accounts for over 98\% of the S$_0$ to S$_1$ character when analyzed with NEO-TDDFT. The progress of the ESIPT reaction can then be tracked by computing the proton position  expectation value as a function of time. The H--\od~and H--\oa~distances are plotted for the two different geometries of \ohba~in Figs. \ref{fig:4panel_weak}C and \ref{fig:4panel_weak}D. In the bare dynamics outside a cavity, the proton of the S$_0$-optimized structure simply relaxes near the donor, with no proton transfer occurring. For the constrained S$_1$-optimized structure, the proton undergoes ultrafast ESIPT, with the crossing point at which the proton is equidistant from \od~and \oa~occurring around 7 fs and the H--\oa~distance minimum occurring around 12 fs.

We now devise a model for a multimode nanocavity that can be coupled to the excited-state proton transfer dynamics of \ohba. To model a lossy, multimode cavity environment, we define the total spectral density $J(\omega)$ with Lorentzian broadening, as in Eq. \ref{eq:lorentzian}, to be
\begin{equation}
J(\omega)=\sum_k J_k(\omega)=\sum_k\frac{g_k^2}{\pi}\frac{\gamma_k/2}{(\omega-\omega_k)^2+(\gamma_k/2)^2}
\label{eq:specdens}
\end{equation}
where each mode has frequency $\omega_k$, light--matter coupling $g_k$, and inverse lifetime $\gamma_k$. Because we only couple to one polarization direction, namely the direction of the S$_0$ to S$_1$ transition dipole vector, we will henceforth simply index modes by $k$ rather than $\alpha$. The multimode structure in this case is modeled as a discretized Gaussian distribution,
\begin{equation}
    g_k=g_0e^{-(\omega_k-\omega_0)^2/\sigma^2}
\end{equation}
where $\omega_0$ and $g_0$ are the frequency and light--matter coupling strength, respectively, at the center of the distribution, and $\sigma^2$ is the variance of the mode distribution. For all \ohba~calculations, $\sigma = 0.25$ eV and $\gamma_k^{-1} = 50$ fs, representing similar parameters to those obtained for plasmonic nanostructures.\cite{Sanchez-Barquilla_Garcia-Vidal_Fernandez-Dominguez_Feist_2022, Medina_Garcia-Vidal_Fernandez-Dominguez_Feist_2021, Cuartero-Gonzalez_Fernandez-Dominguez_2018} 

The Gaussian spectral density is discretized into classical modes that are evenly spaced in energy and extend out to $\pm3\sigma$. \revision{For all simulations in this study, we use 100 classical cavity modes. This number of modes was chosen to provide high resolution for the mode populations, but the dynamics were observed to be well converged with only 50 modes.} All photon modes are initialized with zero momentum and zero displacement. For each \ohba~calculation, $\omega_0$ was set to be the S$_0$ to S$_1$ transition frequency obtained from NEO-TDDFT at that geometry. The corresponding excitation energy, $\Delta E_{\text{S}_1}$, is 3.91 eV for the S$_0$  geometry and 3.29 eV for the constrained S$_1$ geometry. The energy-resolved cavity emission $I_k$ is directly proportional to the mode population, $\langle \hat{a}_k^\dagger \hat{a}_k\rangle = \frac{1}{2\hbar\omega_k}(p^2_k(t) + \omega^2_k q^2_k(t))$, via Eq. \ref{eq:emission}. Therefore, the cavity dynamics can be observed by plotting the mode-specific emission as a function of cavity mode frequency $\omega_k$ and time. For the cavity emission to effectively probe the ultrafast excited-state dynamics, we must assume that cavity loss constitutes the dominant radiative loss channel, as opposed to intrinsic molecular fluorescence. This assumption is supported by the fluorescence lifetime of \ohba, which has been experimentally determined to be on the order of tens of picoseconds, about three orders of magnitude longer than the decay time of lossy nanocavities.\cite{Stock_Bizjak_Lochbrunner_2002}

To explore how a multimode cavity can be used to probe the bare nuclear--electronic dynamics following photoexcitation, we chose a value of $g_0$ that does not disturb these dynamics. For this purpose, we used a spectral density with a light--matter coupling per unit dipole, $g_0/\mu_0$, equal to 0.03 V/nm, where $\mu_0=1$ $e\cdot$nm. This value gives a total semiclassical cavity field strength of $\varepsilon_\text{tot}=(\sum_k \varepsilon_k^2)^{\frac{1}{2}}=1\times10^{-3}$ a.u. \revision{(0.5 V/nm). We report light--matter coupling as $g_0/\mu_0$ because the dipole is a dynamical quantity in real-time simulations. For reference, the fixed S$_0$ to S$_1$ transition dipole moment computed from linear-response NEO-TDDFT is $\sim $ 0.1 $e\cdot$nm ($\sim$5 D), so a coupling of $g_0/\mu_0 = 0.03$ V/nm corresponds to a peak effective coupling of $g_0\sim$3 meV and total coupling of $g_\text{tot}\sim$50 meV.} Figs. \ref{fig:4panel_weak}C and \ref{fig:4panel_weak}D show the proton position expectation value to be identical both inside and outside the cavity, indicating that we are not observing cavity-modified dynamics. The mode-specific cavity emission dynamics are depicted in Figs. \ref{fig:4panel_weak}A and \ref{fig:4panel_weak}B. In both cases, we observe maxima in the emission that are regularly spaced in time with a period of roughly 1.0 fs for the S$_0$ geometry and 1.3 fs for the constrained S$_1$ geometry. These periods correspond roughly to the characteristic timescales of the bare S$_0$ to S$_1$ excitation energies for the two different structures, $\hbar/\Delta E_{\text{S}_1}$, which are 1.06 fs and 1.26 fs, respectively. At early times, the emission is greatest at the vertical S$_0$ to S$_1$ excitation energy, $\Delta E_{\text{S}_1}$. This behavior is expected because these cavity modes are most resonant with the instantaneous transition and have the strongest light--matter coupling $g_k$ due to the spectral density being peaked at this energy. The initial vertical transition energy $\Delta E_{\text{S}_1}$ is marked with a horizontal dashed line on all mode population plots. Shifting of the cavity emission away from this frequency indicates that the nuclear--electronic subsystem has evolved away from the electronic and protonic densities associated with the initial vertical excitation energy.

\begin{figure}
    \centering
    \includegraphics[width=2.9in]{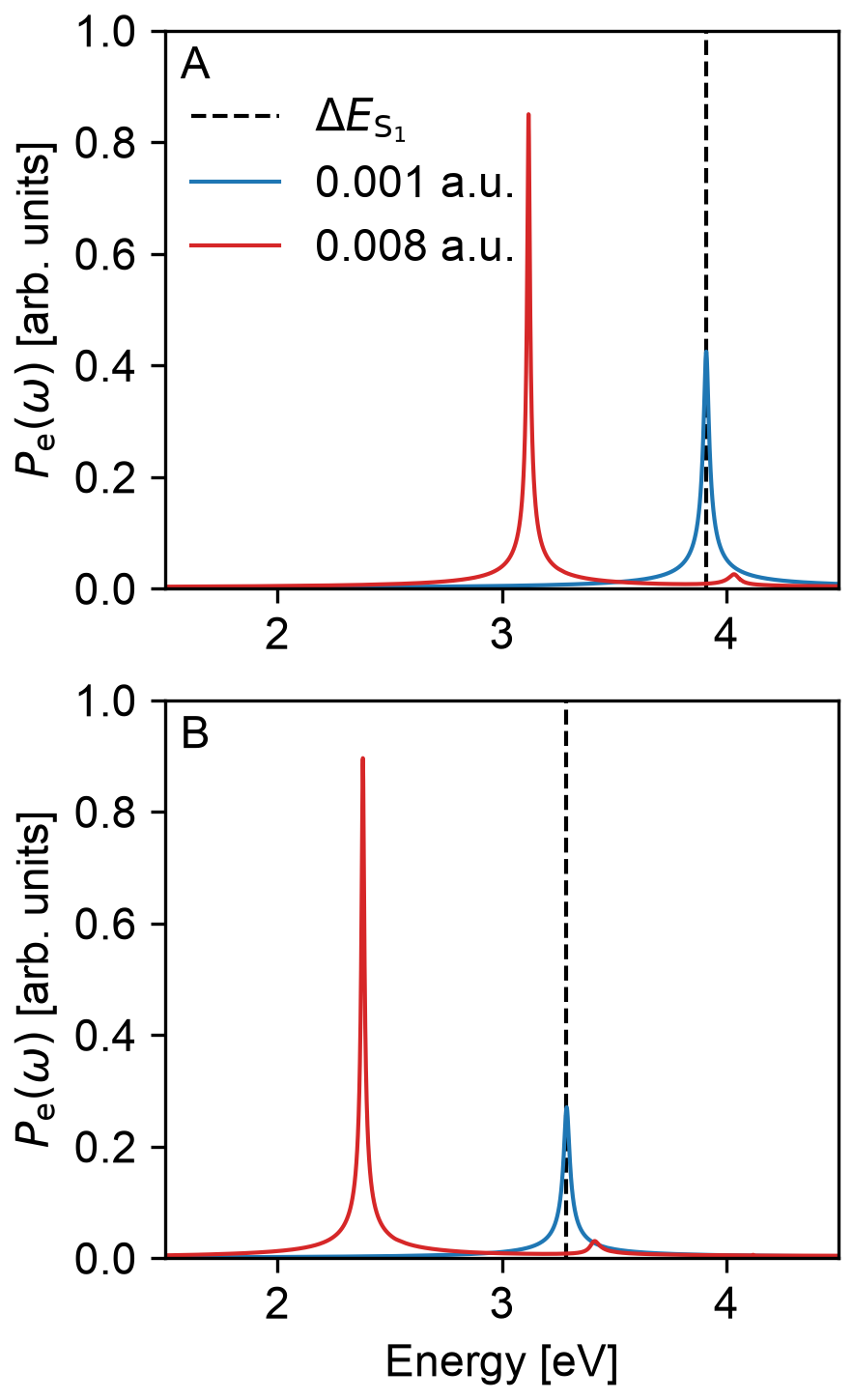}
    \caption{\revision{Electronic power spectrum, $P_\text{e}(\omega)$, of \ohba~coupled to a cavity with a Gaussian spectral density. The spectrum is shown for \ohba~in the (A) S$_0$ geometry and (B) constrained S$_1$ geometry, with total cavity field strength $\varepsilon_\text{tot}=0.001$ a.u. (blue curve) and $\varepsilon_\text{tot}=0.008$ a.u. (red curve). $\Delta E_{\text{S}_1}$ is the NEO-TDDFT vertical S$_0$ to S$_1$ excitation energy of \ohba~at each geometry outside of the cavity, and the cavity spectral density is centered at this frequency.}}
    \label{fig:spectrum_ohba}
\end{figure}

The evolution of the cavity emission maximum away from the initial excitation frequency reflects the changing resonance condition (i.e., excitation energy) for the nuclear--electronic subsystem. This behavior can be observed by tracking the center of the approximately periodic pulses in the two-dimensional map, as well as the oscillations of $\omega^\text{max}_k$, which indicates the frequency of the single mode with the largest cavity emission at each time step.  Using these observables, we can distinguish trajectories where proton transfer does and does not occur. In Fig. \ref{fig:4panel_weak}A, the maximum cavity emission is predominantly centered around $\Delta E_{\text{S}_1}$. The resonance finally shifts to a slightly lower energy at around 9 -- 10 fs, indicating relaxation of the proton near the donor on the S$_1$ surface. For the ultrafast proton transfer in Fig. \ref{fig:4panel_weak}B, the shift to a lower resonant energy occurs much more dramatically and rapidly. This shift in the resonance is due to the proton evolution on the S$_1$ surface toward the acceptor, corresponding to a much lower instantaneous S$_0$ to S$_1$ transition energy (Fig. \ref{fig:ohba_pes}B). \revision{The observation that this shift toward lower-frequency cavity modes coincides with the proton transfer timescale and only occurs in the restricted S$_1$ geometry, where the proton readily transfers, is strong evidence that this shift is a signature of ESIPT rather than some other relaxation process.} It is important to note that the new maximum in the cavity emission no longer coincides with the maximum light--matter coupling, which is peaked at $\Delta E_{\text{S}_1}$. Nevertheless, the intensity of the cavity emission, and therefore the magnitudes of the associated mode populations, increases dramatically after proton transfer occurs. This behavior indicates that the proton density has become more localized near the acceptor, corresponding to a narrower distribution of excitation energies that are now strongly resonant with the lower-energy cavity modes.
\begin{figure*}[htp]
    \centering
    \includegraphics[width=6in]{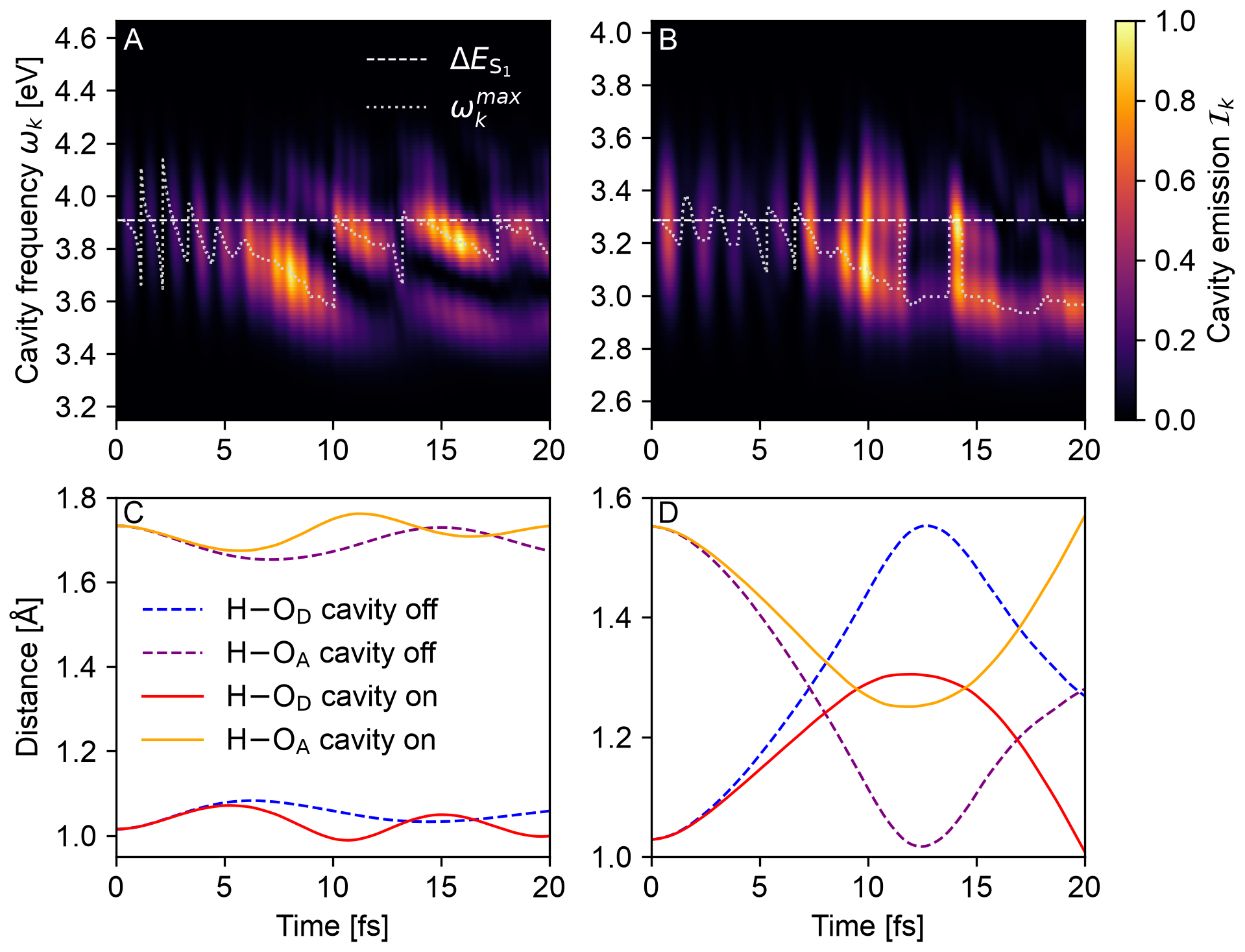}
    \caption{Upper plots: Mode-specific cavity emission, $I_k$, indicated by color intensity, as a function of mode frequency $\omega_k$ and time, for \ohba~in the (A) S$_0$ geometry and (B) constrained S$_1$ geometry. The cavity environment is represented by a Gaussian spectral density peaked at the initial NEO-TDDFT vertical S$_0$ to S$_1$ excitation energy, $\Delta E_{\text{S}_1}$, with $g_0/\mu_0=0.24$ V/nm and $\sigma=0.25$ eV. The frequency of the cavity mode with the largest emission as a function of time is $\omega^\text{max}_k$. Emission is normalized by the single largest intensity observed across the entire time series.  Lower plots: H--\od~and H--\oa~distances in \AA ngstroms as a function of time for \ohba~in the (C) S$_0$ geometry and (D) constrained S$_1$ geometry. The distances are computed using the proton position expectation value. The results are shown with the cavity off \revision{(blue and violet dashed) and on (red and orange solid)}.}
    \label{fig:4panel_strong}
\end{figure*}

After demonstrating that emission from a multimode nanocavity can passively report on ultrafast nuclear--electronic dynamics, we now explore the regime where the cavity actively modifies the dynamics. For this purpose, we use the same Gaussian spectral density but increase the total light--matter coupling per unit dipole by a factor of eight to $g_0/\mu_0=0.24$ V/nm, corresponding to a total semiclassical field strength of $\varepsilon_\text{tot}=8\times 10^{-3}$ a.u. \revision{(4 V/nm)}. \revision{To confirm that these parameters lead to strong light--matter coupling, we analyze the electronic power spectrum, 
\begin{equation}
P_\text{e}(\omega)=\sum_{s=x,y,z}\mathcal{F}[\langle\mathbf{\mu^\text{e}_s(t)}\rangle]
\end{equation}
where $\mathcal{F}$ indicates the Fourier transform and $\mu^\text{e}_s(t)$ is the electronic dipole moment in the $s$ direction. The dipole signal was simulated for 100 fs, and the Fourier transform was performed with the Pad\'e approximation\cite{Bruner_LaMaster_Lopata_2016,Goings_Lestrange_Li_2018} to improve the spectral resolution for the relatively short simulation time. In contrast to the photoexcitation process used to simulate ESIPT, the molecule was started in the NEO-DFT ground state at the S$_0$ or constrained S$_1$ geometry, and a delta pulse at $t=0$ with an electric field amplitude of 0.01 a.u. in the direction along the S$_0$ to S$_1$ transition dipole vector was applied to the cavity modes.}

\revision{The electronic power spectrum of \ohba, $P_\text{e}(\omega)$, is shown in Fig. \ref{fig:spectrum_ohba} with $\varepsilon_\text{tot}=0.001$ a.u. and $\varepsilon_\text{tot}=0.008$ a.u. for both the S$_0$ (Fig. \ref{fig:spectrum_ohba}A) and constrained S$_1$ (Fig. \ref{fig:spectrum_ohba}B) geometries. These simulations confirm that in the intermediate coupling regime with $\varepsilon_\text{tot}=0.001$ a.u., no Rabi splitting is present, and the only effect of the cavity is a natural Lorentzian linewidth due to cavity loss. For the case with $\varepsilon_\text{tot}=0.008$ a.u., however, we observe a pronounced Rabi splitting of $\sim$ 0.8--0.9 eV, corresponding to Rabi oscillations with a period of $\sim$ 4--5 fs. Both the splitting and intensity are asymmetric, with a greater peak shift and intensity for the lower polariton peak. This asymmetry has previously been observed in RT-NEO calculations of oHBA,\cite{Li_Tao_Hammes-Schiffer_2022} as well as QED electronic structure calculations,\cite{Flick_Narang_2020,Pavosevic_Flick_2021} and is associated with coupling to higher-lying excited states.\cite{Li_Hammes-Schiffer_2023a,Yang_Ou_Pei_Wang_etal_2021} Nonetheless, the splitting of the molecular transition peak into upper and lower polariton peaks is a clear signature of a hybrid light--matter state due to strong coupling.}

\revision{In addition to Rabi splitting in the power spectrum, strong light--matter coupling can lead to changes in the proton transfer dynamics.\cite{Li_Tao_Hammes-Schiffer_2022}} Figs. \ref{fig:4panel_strong}C and \ref{fig:4panel_strong}D track the proton transfer dynamics via the H--\od~and H--\oa~distances. In both cases, the dynamics are modified by the presence of the cavity. The most notable effect of the cavity is observed for the constrained S$_1$ geometry in Fig. \ref{fig:4panel_strong}D, where the proton is inhibited from fully transferring before recrossing back to the donor side. In this regime, the cavity mode populations and corresponding cavity emission (Figs. \ref{fig:4panel_strong}A and B) show the time-resolved onset of strong coupling. For the first several fs, the cavity emission behaves very similarly to the intermediate coupling case. During this time, the oscillations in emission intensity are dictated solely by the bare nuclear--electronic dynamics. At around 7 fs, the dynamics become more complex with multiple timescales appearing. These dynamics are characterized by rapid modulations of the overall emission intensity and longer timescale, Rabi-like oscillations of the mode frequency at which maximum emission occurs. \revision{Although the oscillations are not regular enough to assign a precise period, the emission maxima appear at an interval of $\sim$ 4-5 fs, roughly matching the timescale associated with the Rabi splitting in Fig. \ref{fig:spectrum_ohba}.}
\begin{figure*}
    \centering
    \includegraphics[width=6in]{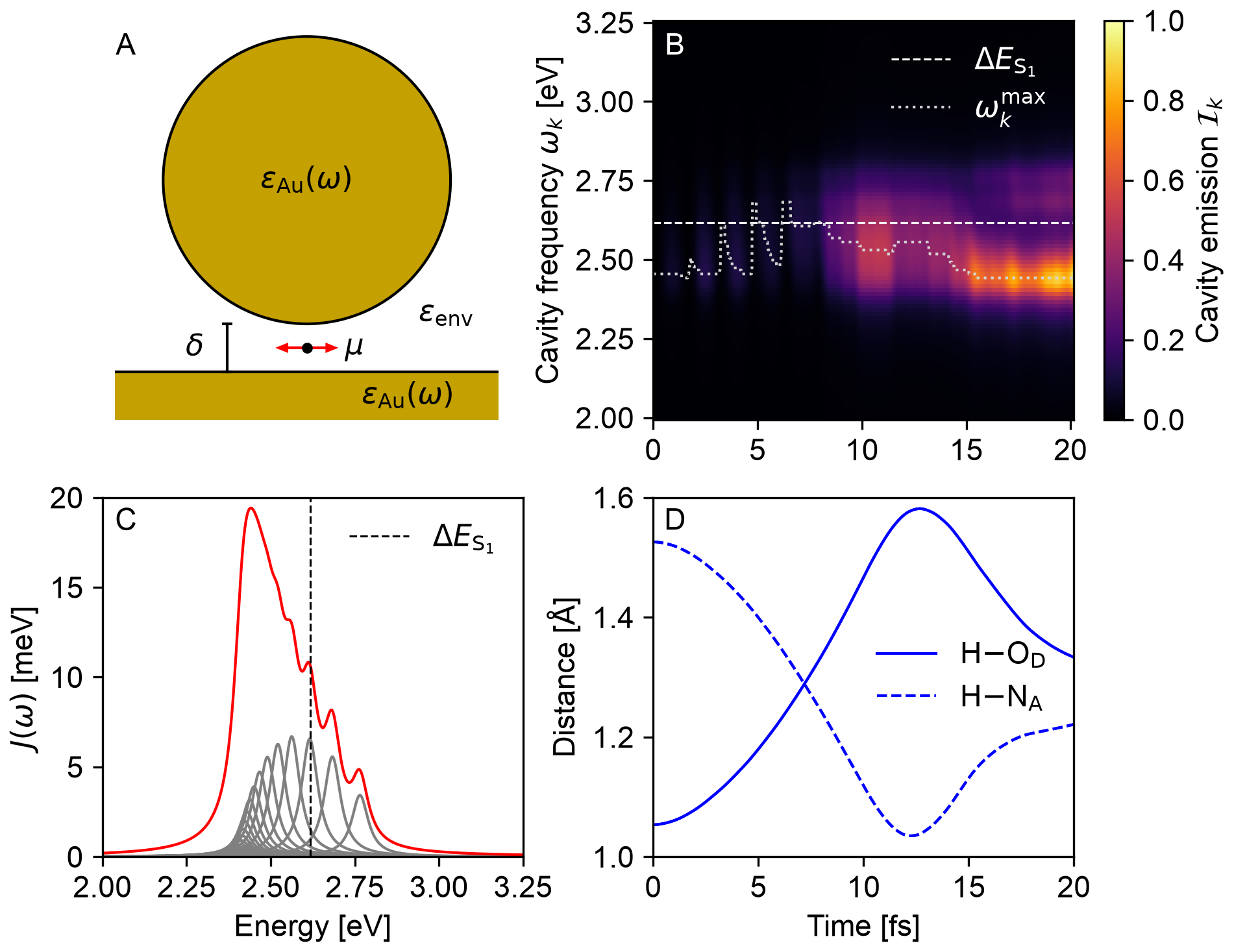}
    \caption{(A) Schematic depiction of gold nanoparticle-on-mirror setup (not to scale). $\epsilon_\text{Au}(\omega)$ is the dielectric function for gold, $\epsilon_\text{env}$ is the dielectric constant of the solvent environment, $\delta$ is the distance between the nanoparticle and the mirror, and $\mu$ is the emitter dipole. (B) Mode-specific cavity emission, $I_k$, indicated by color intensity, as a function of mode frequency $\omega_k$ and time, for AMIEP in the constrained S$_1$ geometry. The frequency of the cavity mode with the largest emission as a function of time is $\omega^\text{max}_k$. Emission is normalized by the single largest intensity observed across the entire time series.  $\Delta E_{\text{S}_1}$ is the initial NEO-TDDFT vertical S$_0$ to S$_1$ excitation energy of AMIEP. (C) Spectral density $J(\omega)$ for the NPoM setup depicted in (A). The emitter dipole is oriented parallel to the mirror at $\delta/2$, and the spectral density is plotted with a dipole value of 1 $e\cdot$nm and $\epsilon_\text{env}=4$. Each individual Lorentzian mode is shown in gray, while the total summation from Eq. \ref{eq:specdens} is shown in red. (D) H--\od~and H--N$_\text{A}$ distances in \AA ngstroms as a function of time for AMIEP in the constrained S$_1$ geometry. The distances are computed using the proton position expectation value. The results are shown with the cavity off (blue), but the curves are indistinguishable with the cavity on.}
    \label{fig:npom}
\end{figure*}

\subsection*{Excited-state proton transfer in a  \revision{model} gold nanoparticle-on-mirror cavity}
Heretofore, our results have relied on a tunable Gaussian spectral density to represent the nanocavity. The smooth distribution and dense, evenly-spaced discretization enable high resolution with respect to cavity mode frequency. To support our conclusion that a multimode nanocavity emission could be an experimentally viable probe of nuclear--electronic quantum dynamics, we now focus on modeling an experimentally relevant example. The nanoparticle-on-mirror (NPoM) experimental configuration has been used to achieve strong coupling at the molecular level\cite{Chikkaraddy_deNijs_Benz_Barrow_Scherman_Rosta_etal_2016} as well as for enhanced spectroscopy applications.\cite{Peng_Zhou_Li_Sun_etal_2024,Benz_Chikkaraddy_Salmon_Ohadi_etal_2016} Furthermore, the spectral density of a dipole emitter in a NPoM cavity is analytically solvable via the transformation optics approach.\cite{Pendry_Aubry_Smith_Maier_2012,Cuartero-Gonzalez_Fernandez-Dominguez_2018,Cuartero-Gonzalez_2020} The solution for the spectral density gives the same multimodal Lorentzian form as Eq. \ref{eq:specdens}, with mode-specific frequencies $\omega_k$, effective light--matter couplings
$g_k$, and inverse lifetimes $\gamma_k$. Details for computing $J(\omega)$ for a given NPoM geometry and dielectric function are provided in Ref. \citenum{Cuartero-Gonzalez_Fernandez-Dominguez_2018}.

We compute the spectral density \revision{for a model} gold NPoM using the material parameters previously devised\cite{Cuartero-Gonzalez_Fernandez-Dominguez_2018} to emulate the experimental conditions in Ref. \citenum{Chikkaraddy_deNijs_Benz_Barrow_Scherman_Rosta_etal_2016}. \revision{
The gold nanoparticle and mirror are represented with a Drude dielectric function $\epsilon_\text{Au}(\omega)$: 
\begin{equation}
\epsilon_\text{Au}(\omega)=\epsilon_\infty - \frac{\omega_\text{p}^2}{\omega(\omega-i\gamma)}
\end{equation}
where $\omega_\text{p}=8.91$ eV is the plasmon frequency, $\epsilon_\infty=9.7$ is the high-frequency dielectric constant, and $\gamma=0.06$ eV is the plasmon inverse lifetime. The nanoparticle radius is 30 nm, the separation from the mirror is $\delta = 0.9$ nm, and the environment dielectric constant is $\epsilon_\text{env} =  4.0$. This dielectric continuum representation of the cavity can model electrodynamic effects that lead to light--matter coupling, but other relevant effects, such as electron transfer or surface interactions, require an atomistic cavity description and are therefore not captured. Although metal clusters can be explicitly modeled at an atomistic level with RT-NEO methods,\cite{Li_Hammes-Schiffer_2023,Li_Paenurk_Hammes-Schiffer_2024} a gold NPoM cavity would require much larger metal clusters that are currently inaccessible.}  We orient the emitter dipole parallel to the mirror plane at $\delta/2$, which produces a smooth mode distribution with coupling strengths in the intermediate-coupling regime. The nanocavity setup and spectral density are shown in Figs. \ref{fig:npom}A and \ref{fig:npom}C, respectively. 

\begin{figure}[htp]
    \centering
    \includegraphics[width=3.0in]{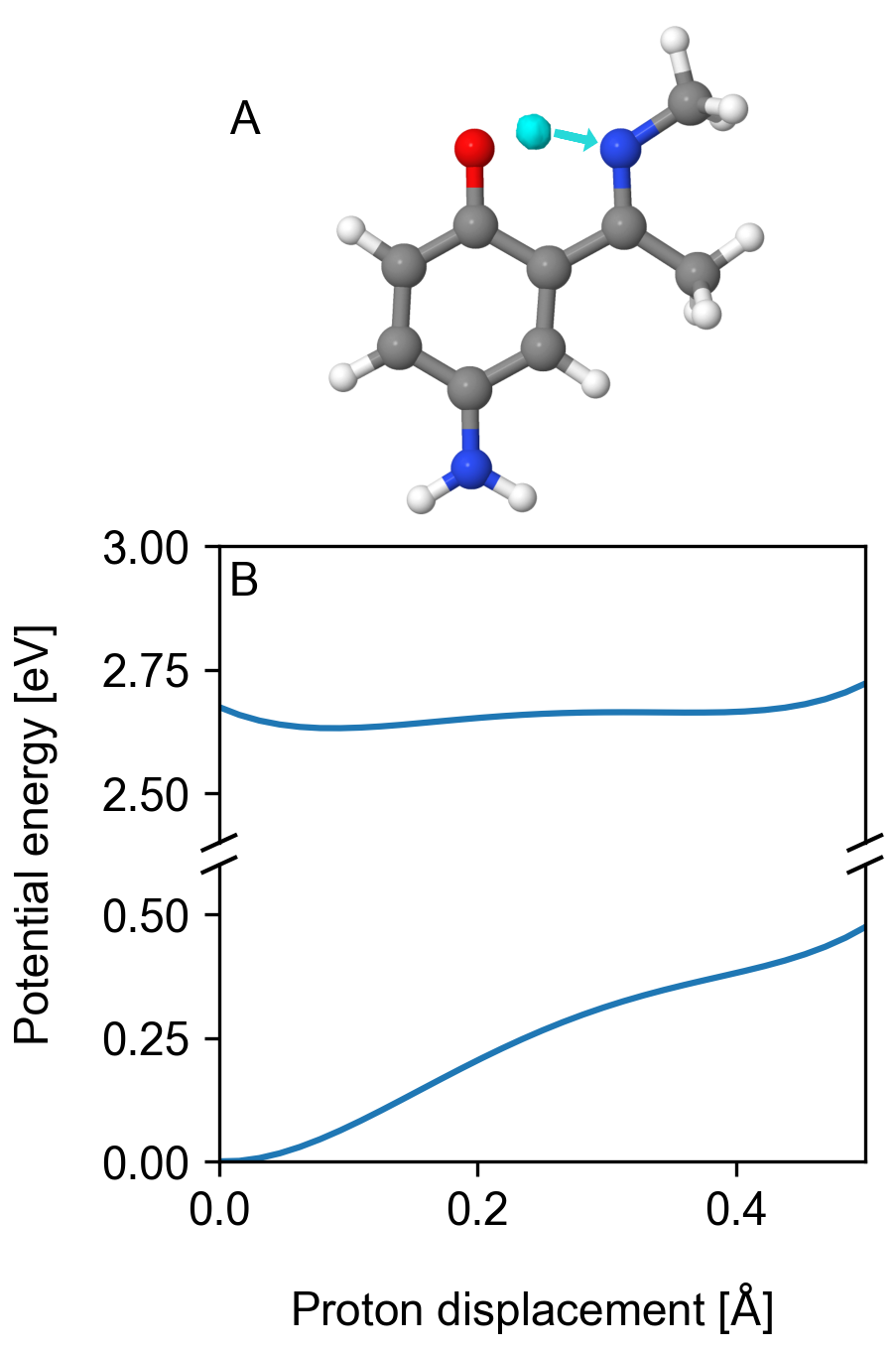}
    \caption{(A) Structure of the Schiff base, 4-amino-2-[(1-(methylimino)ethyl)]phenol (AMIEP), in the constrained S$_1$ geometry. The isosurface of the transferring proton density computed with NEO-DFT is shown in cyan. The cyan arrow depicts the proton transfer coordinate used for computing the potential energy profiles. (B) S$_0$ and S$_1$ proton potential energy profiles for AMIEP at the constrained S$_1$ geometry, as computed with conventional TDDFT by scanning the proton coordinate along the axis connecting its optimized ground state position to N$_\text{A}$.}
    \label{fig:schiff_pes}
\end{figure}

Typical NPoM setups produce modes in the 2 -- 3 eV range and thus would require modification of the material parameters to be in resonance with the S$_0$ to S$_1$ transition of \ohba. In order to maintain a faithful representation of our nanocavity setup, we instead found a different small molecule that will undergo ESIPT with lower excitation energies. Specifically, we will use the Schiff base 4-amino-2-[1-(methylimino)ethyl]phenol (AMIEP). Using conventional TDDFT, we computed the  S$_0$ to S$_1$ excitation energy to be 2.61 eV for the constrained S$_1$ geometry, which was optimized using the same protocol as discussed above for \ohba. The structure of AMIEP is shown in Fig. \ref{fig:schiff_pes}A with the protonic density computed via NEO-DFT. The proton potential energy profile (Fig. \ref{fig:schiff_pes}B) exhibits both an overall lower S$_0$ to S$_1$ excitation energy and a lower proton transfer barrier than \ohba. \revision{Furthermore, the proton transfer dynamics are very similar between oHBA and AMIEP. The AMIEP proton dynamics are shown in Fig. 5D. For both oHBA and AMIEP in the constrained S$_1$ geometry, the transferring proton becomes equidistant between the donor and acceptor at around 7.5 fs, and the closest approach to the acceptor occurs at around 12.5 fs. We note that AMIEP will also undergo proton transfer in the S$_0$ optimized geometry; however, because the excitation energy is much higher, this configuration is not useful for coupling to the NPoM cavity.}

We now examine the proton transfer dynamics of AMIEP coupled to a realistic NPoM cavity. Fig. \ref{fig:npom}C shows the S$_0$ to S$_1$ vertical excitation energy relative to the NPoM spectral density. Although this excitation energy overlaps with some high-lying cavity modes, the strongest overall light--matter coupling and greatest density of cavity modes occurs at 2.41 eV. Fig. \ref{fig:npom}D shows the H--\od~and H--N$_\text{A}$ distances as a function of time, indicating that proton transfer indeed occurs for AMIEP on the ultrafast timescale. We plot only the out-of-cavity H--\od~and H--N$_\text{A}$ distances because the in-cavity dynamics are virtually identical. Turning our attention to the mode-specific cavity emission, which is shown in Fig. \ref{fig:npom}B, we find that the intensity is very low at early times. In contrast to the model Gaussian spectral densities, which were peaked at the \ohba~S$_0$ to S$_1$ vertical transition, the NPoM has very few cavity modes near the vertical transition for AMIEP. Furthermore, although the modes were evenly distributed in the Gaussian discretization, for the NPoM spectral density they are very sparse at higher frequencies and are heavily concentrated near 2.4 eV. Because of this energetic mismatch, the modes remain largely unpopulated, and consequently there is very little cavity emission for the first 7 fs. When the nuclear--electronic dynamics finally evolve toward a state corresponding to a lower excitation energy at around 10 fs, the modes in the middle of the distribution, near 2.5 eV, begin to become populated, gaining noticeable emission intensity. Finally, at around 15 fs, the main peak accounts for the dominant cavity emission because it has the strongest light--matter coupling, and the molecular subsystem has evolved to a state with a low enough excitation energy to be in resonance. Thus, we have shown that the evolution of the nuclear--electronic subsystem is reflected by a changing resonance condition, and consequently it can be observed in the changing cavity emission. 

In this cavity setup, there are no signs of strong coupling, such as modified proton dynamics or Rabi-like features. Reaching the strong-coupling limit requires fine tuning of the cavity to be in precise resonance with the molecule. For the NPoM/AMIEP system, we tune the physically accessible parameters to deduce if polariton formation is feasible. To monitor this possibility, we analyze the electronic power spectrum, $P_\text{e}(\omega)$. All parameters for the calculation of $P_\text{e}(\omega)$ are identical to those used to generate Fig. \ref{fig:spectrum_ohba}.

\begin{figure}[htp!]
    \centering
    \includegraphics[width=2.9in]{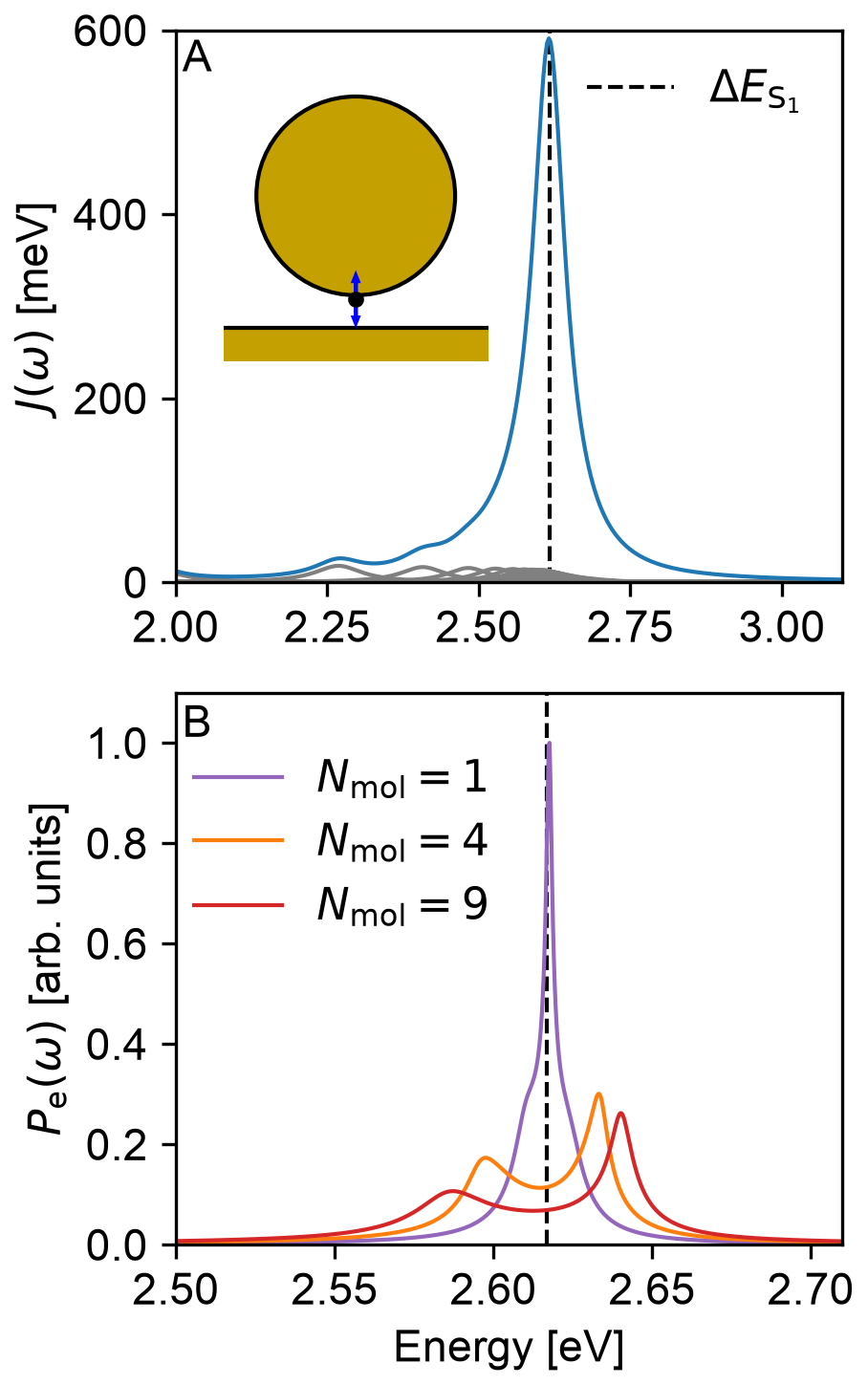}
    \caption{(A) Spectral density $J(\omega)$ for the NPoM setup with the emitter dipole oriented perpendicular to the mirror at $7\delta/8$, plotted with a dipole value of 1 $e\cdot$nm and $\epsilon_\text{env}=1.9$. The cavity and dipole orientation are depicted in the inset. Each individual Lorentzian mode is shown in gray, while the total summation from Eq. \ref{eq:specdens} is shown in blue. $\Delta E_{\text{S}_1}$ is the initial NEO-TDDFT vertical S$_0$ to S$_1$ excitation energy of AMIEP. (B) Electronic power spectrum, $P_\text{e}(\omega)$, of AMIEP coupled to the NPoM spectral density. To model the effect of multiple molecules in the cavity, the total light--matter coupling is scaled by $\sqrt{N_\text{mol}}$ with $N_\text{mol}=1$, 4, and 9. To show the details more clearly, the energy range is narrower in panel (B) than in panel (A).}
    \label{fig:spectrum_amiep}
\end{figure}

To move toward the strong-coupling regime, we must increase the overall coupling and move the main spectral peak into resonance with the electronic S$_0$ to S$_1$ vertical transition. To increase the coupling, we align the emitter dipole perpendicular to the mirror and move it closer to the nanoparticle surface at $7\delta/8$. This displacement is meant to model the molecule on the nanoparticle surface but could lead to steric crowding. Although geometric parameters can affect the overall coupling, the main peak location is determined by the dielectric properties $\epsilon_\text{env}$ and $\epsilon_\text{Au}(\omega)$. Because the environmental dielectric constant is the easiest to modify experimentally by changing the solvent, we tune the frequency with $\epsilon_\text{env}$ and leave $\epsilon_\text{Au}(\omega)$ unchanged. A value of $\epsilon_\text{env}=1.9$, which is similar to that of a nonpolar solvent such as $n$-hexane, brings the main peak of the spectral density to 2.62 eV,  nearly resonant with the AMIEP S$_0$ to S$_1$ transition of 2.61 eV. The resulting spectral density compared to the electronic transition frequency, as well as a depiction of the cavity setup, is shown in Fig. \ref{fig:spectrum_amiep}A. We now examine the resulting electronic power spectrum for polariton formation in the form of a Rabi splitting.

Even at resonance, the NPoM cavity coupled to a single molecule does not produce well-defined upper and lower polariton peaks. The electronic power spectrum for the AMIEP single molecule is given as the purple trace in Fig. \ref{fig:spectrum_amiep}B. The main peak is shifted up in energy compared to the bare excitation, while a shoulder appears below, which could indicate Rabi splitting that is poorly resolved due to the linewidth of the NPoM spectral density. We note that the experimental single-molecule strong coupling setup used organic dye molecules as the emitters. Such dye molecules have much larger transition dipole moments than the molecules studied here and thus will show stronger coupling.\cite{Chikkaraddy_deNijs_Benz_Barrow_Scherman_Rosta_etal_2016} Additionally, in the experiments, several emitters could aggregate in each cavity site, and single-molecule spectra had to be extracted from data within a statistical ensemble of number of molecules, $N_\text{mol}$, per cavity. Assuming that the emitters independently couple to the field and not to one another, the overall light--matter coupling will be scaled by $\sqrt{N_\text{mol}}$ in each cavity. With this in mind, we scaled the light--matter coupling by $\sqrt{N_\text{mol}}=2$ and 3, corresponding to $N_\text{mol}=4$ and 9, respectively. For both $N_\text{mol}=4$ and $N_\text{mol}=9$, clear upper and lower polariton peaks can be observed with a Rabi splitting around 40 meV and 60 meV, respectively. These results demonstrate that strong coupling can be theoretically observed with RT-NEO-TDDFT using an experimentally relevant nanocavity setup and number of molecules per cavity.

\section*{Conclusions}
In this work, we used RT-NEO-TDDFT with classical cavity modes to study the effects of a lossy, multimode cavity environment on excited-state proton transfer. This type of electromagnetic environment is typical of sub-wavelength plasmonic nanocavities that are used for single- or few-molecule strong coupling and enhanced spectroscopy applications. Focusing on ESIPT in \ohba, we show that the energy- and time-resolved cavity emission can be used to track bare nuclear--electronic quantum dynamics in the intermediate-coupling regime. In this case, proton transfer is reflected in the cavity emission evolving toward lower-energy modes. In the strong-coupling regime, after an initial relaxation similar to the bare dynamics, an onset of Rabi-like oscillations along with a suppression of proton transfer due to polariton formation is observed. We also applied this technique with an experimentally relevant spectral density corresponding to a nanoparticle-on-mirror cavity for a different ESIPT molecule. In this case, we found that even when the dominant cavity mode is not resonant with the initial excitation, a molecule undergoing ESIPT will become resonant as relaxation occurs. When the dominant cavity peak is tuned to be resonant with the electronic transition, polariton formation cannot be clearly observed for a single molecule but is observed for a small collection of molecules.

\revision{Although the molecules and cavity setups used here serve as illustrative examples, the underlying principles are broadly applicable. Many molecules that undergo ESIPT have S$_0$ to S$_1$ transitions in the 2--4 eV range, which is precisely the range covered by many nanocavity setups. This study has demonstrated that a cavity whose spectral density is sharply peaked at resonance with this transition is most likely to produce signatures of strong coupling, such as Rabi splitting and proton transfer suppression, especially when the molecule has a large S$_0$ to S$_1$ transition dipole. In contrast, cavities with broad, strongly multimodal spectral densities are useful as dynamical probes via cavity emission.  However, such cavities are most useful when paired with molecules that undergo proton transfer on a timescale that is shorter than or comparable to the cavity lifetime.}

The RT-NEO framework with classical cavity modes is a powerful platform for studying nuclear--electronic quantum dynamics in complex electromagnetic environments.  \revision{The results suggest that far-field cavity emission could be an experimental observable for probing the timescale of proton transfer reactions. This framework could also be used to study hydrogen/deuterium isotope effects with cavity emission as a means for resolving the proton transfer mechanism.} In the future, such multimode setups could be formulated to couple with multiple transitions simultaneously,\cite{Ke_Assan_2025} which could include both vibrational and electronic transitions in the NEO framework. In general, this framework allows us to move toward the simulation of physically realistic and experimentally attainable cavity environments.

\begin{acknowledgement}
The authors thank Millan Welman, Scott Garner, and Arghadip Koner for helpful discussions. 
\end{acknowledgement}

\section*{Funding Sources}
This material is based upon work supported by the Air Force Office of Scientific Research under AFOSR Award No. FA9550-24-1-0347.

\begin{suppinfo}
\begin{description}
\item \revision{Summary of cavity parameters used for generating Gaussian and nanoparticle-on-mirror spectral densities (PDF).}
\end{description}
\end{suppinfo}

%\section*{Associated Content}
%J. H. Fetherolf; Tim Duong; T. E. Li; S. Hammes-Schiffer. Nuclear--Electronic Quantum Dynamics in a Plasmonic Nanocavity. 2026, arXiv:2603.12373v2. arXiv. \url{https://arxiv.org/abs/2603.12373v2} (accessed June 19, 2026).

\section*{Data Availability Statement}
\revision{The data that support the findings of this study will be openly available upon publication at the Zenodo repository with digital object identifier \url{http://doi.org/10.5281/zenodo.20174149}.}

%\begin{suppinfo}
%\end{suppinfo}
\bibliography{refs}

\clearpage
\section*{For Table of Contents Use Only}

\noindent\textbf{Nuclear--Electronic Quantum Dynamics in a Plasmonic Nanocavity}

\vspace{0.3em}
\noindent Jonathan H. Fetherolf, Tim Duong, Tao E. Li, and Sharon Hammes-Schiffer

\vspace{1em}
\begin{center}
    \includegraphics[width=3.25in]{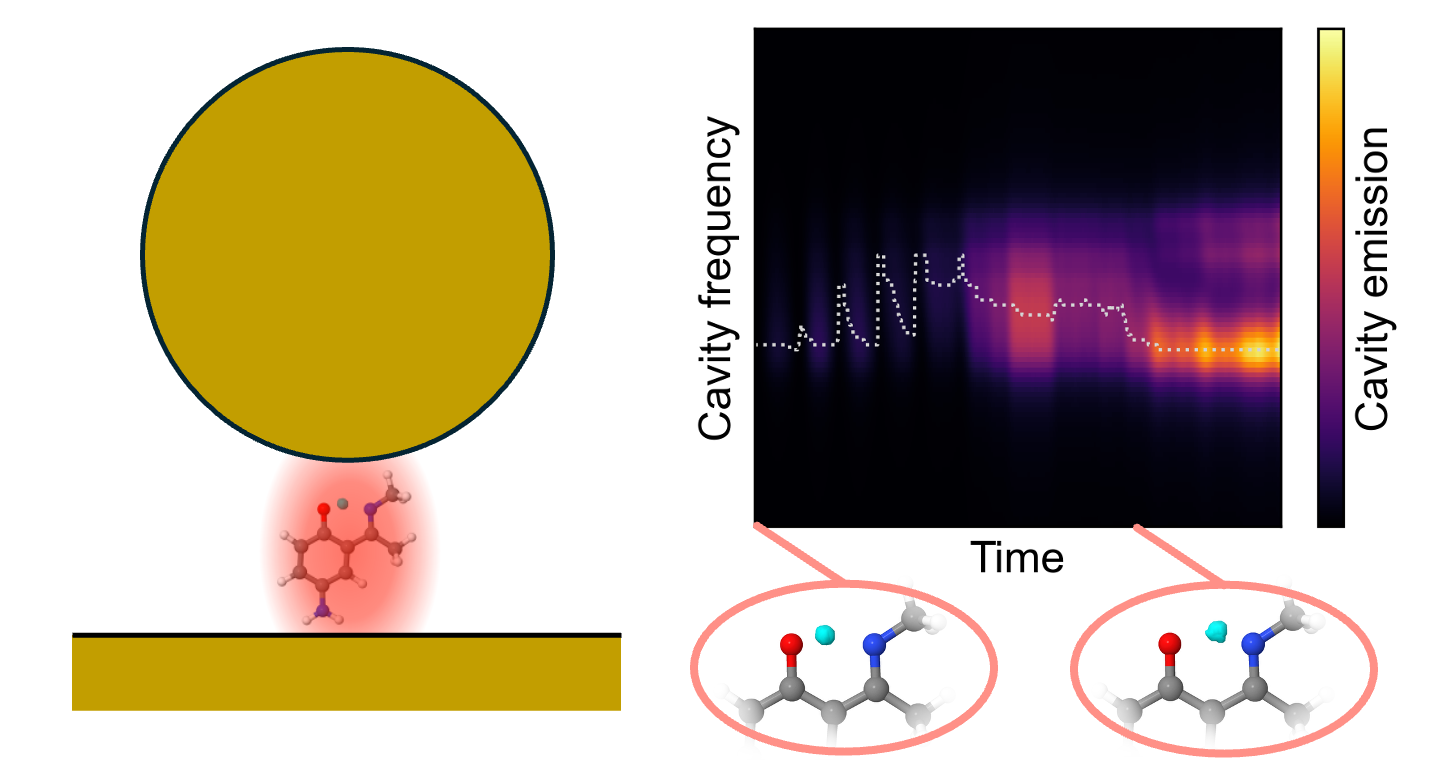}
\end{center}

\vspace{0.5em}
\noindent Left: Depiction of gold nanoparticle-on-mirror nanocavity containing a proton transfer molecule. Right: Frequency-resolved cavity emission as a function of time with snapshots of proton transfer occurring.

\includepdf[pages={1,2,3}]{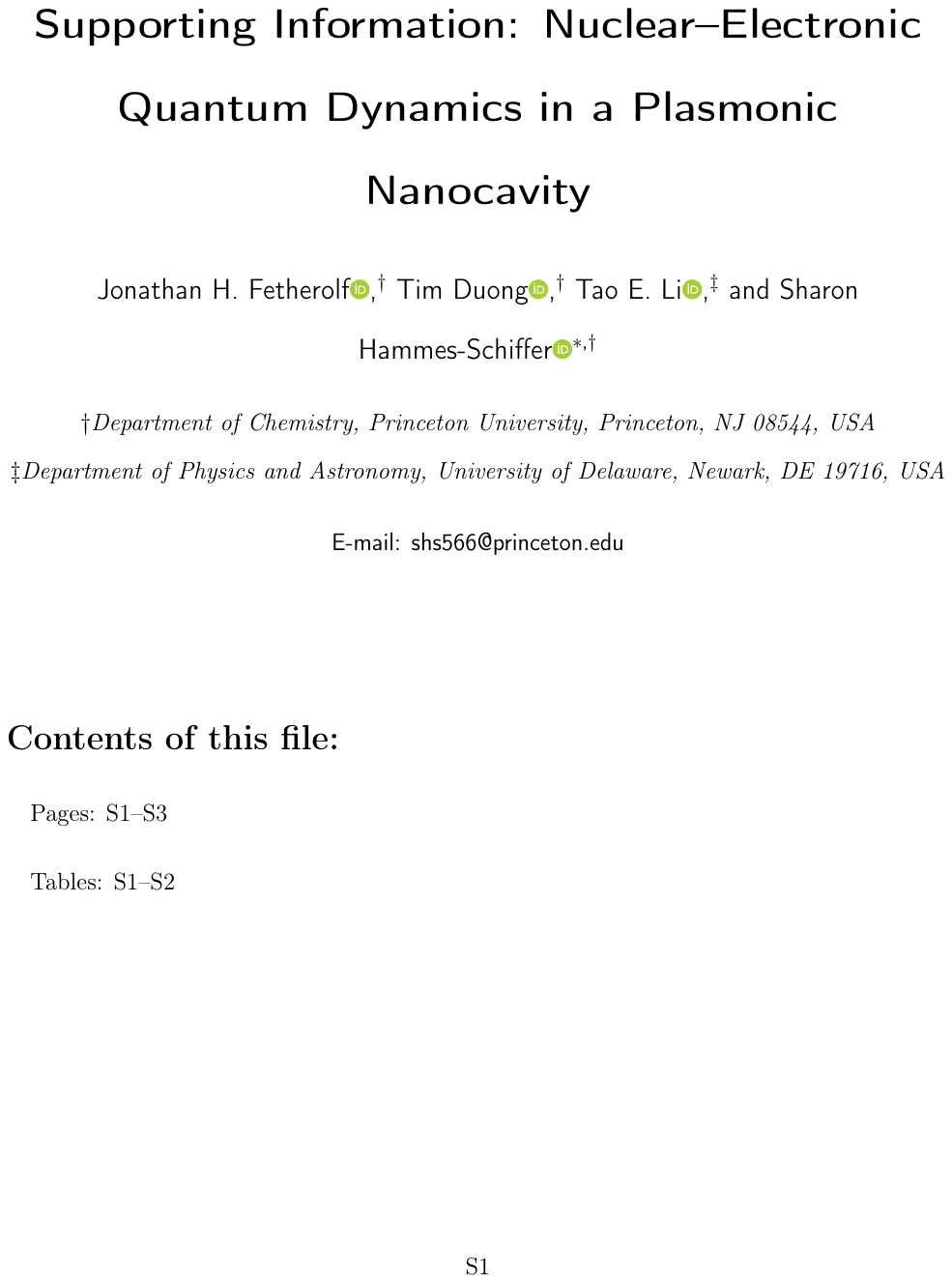}
\end{document}